\documentclass[twocolumn,  tighten]{aastex62}
\usepackage{natbib}

\def\MassiveFIRE{{\sc MassiveFIRE}}
\def\MUSIC{{\sc MUSIC}}

\submitjournal{The Astrophysical Journal Letters}

\shorttitle{The galaxy -- halo connection in low mass halos}
\shortauthors{Feldmann et al.}

\begin{document}

\title{The galaxy -- halo connection in low mass halos}

\correspondingauthor{Robert Feldmann}
\email{feldmann@physik.uzh.ch}

\author[0000-0002-1109-1919]{Robert Feldmann}
\affil{Institute for Computational Science, \\
University of Zurich, \\
Zurich CH-8057, Switzerland}

\author[0000-0002-4900-6628]{Claude-Andr\'{e} Faucher-Gigu\`{e}re}
\affil{Department of Physics and Astronomy and CIERA, \\
Northwestern University, \\
Evanston, IL 60208, USA}

\author[0000-0002-1666-7067]{Du\v{s}an Kere\v{s}}
\affil{Center for Astrophysics and Space Sciences, \\
University of California San Diego, \\
San Diego, CA 92093, USA}

\begin{abstract}
Properties of galaxies vary systematically with the mass of their parent dark matter halos. This basic galaxy -- halo connection shows a fair amount of scatter whose origin is not fully understood. Here, we study how differences in the halo assembly history affect central galaxies in low mass ($M_{\rm halo}<10^{12}$ $M_\odot$) halos at $z=2-6$ with the help of the \MassiveFIRE{} suite of cosmological simulations. In contrast to previous works that tie galaxy properties to halo concentration and halo formation redshift, we focus on halo growth rate as a measure of assembly history. We find that, at fixed halo mass, faster growing halos tend to have lower stellar masses and higher SFRs per unit stellar mass but similar overall SFRs. We provide a simple explanation for these findings with the help of an analytic model that captures approximately the behavior of our hydrodynamical simulations. Specifically, among halos of a given current mass, quickly growing halos have lower stellar masses (and thus higher sSFRs) because they were less massive and had comparably lower cold gas masses and SFRs in the past than slowly growing halos. By combining these findings with estimates for the scatter of the halo growth rate, we show that variations in growth rate at fixed halo mass may largely explain the scatter of the stellar mass -- halo mass relation. In contrast, halo growth variations likely play only a minor role in the scatter of the star forming sequence in low mass galaxies.
\end{abstract}


\section{Introduction}

The evolution of galaxies is tied to the evolution of the dark matter halos in which they reside \citep{White1978}. Empirical models developed over the past decade successfully utilize this galaxy -- halo connection by linking galaxy properties and the masses of dark matter halos. For instance, sub-halo abundance matching \citep{Kravtsov2004, Behroozi2013c} and halo occupation distribution models \citep{Peacock2000, Kravtsov2004}, which parametrize galaxy properties by halo mass alone, correctly predict the clustering of galaxies of different masses and luminosities.
However, mass is only one of many properties of dark matter halos, although arguably the most fundamental. In fact, at fixed halo mass, halo clustering has been shown to depend on halo properties related to the halo assembly such as halo formation time \citep{Gao2005}, halo concentration \citep{Wechsler2006}, or sub-halo fraction \citep{Zhu2006}. This result, known as assembly bias, raises the question of whether or not galaxy properties also depend on the detailed assembly histories of halos of a given mass. 

Semi-analytic models (e.g., \citealt{Croton2007a}), semi-empirical models (e.g., \citealt{Moster2018}), and hydrodynamical simulations (e.g., \citealt{Faucher-Giguere2011, Feldmann2015, Feldmann2016, Feldmann2017, Chaves-Montero2016, Matthee2017}) predict that stellar masses and specific star formation rates (sSFRs) of galaxies are correlated with the growth histories of their halos. Observationally, the situation is less clear, with some studies claiming detection \citep{Wang2013a, Tinker2016a} while others do not (e.g., \citealt{Blanton2007, Lin2015}).

Ignoring this `galaxy assembly bias' may systematically affect the modeling of the galaxy -- halo connection \citep{Zentner2014}. Furthermore, galaxy assembly bias is potentially an important contributor to the observed scatter in galaxy scaling relations such as the stellar mass -- halo mass relation (SHMR; e.g., \citealt{Matthee2017}) or the SFR -- stellar mass relation (e.g., \citealt{Rodriguez-Puebla2016}). Hence, deciphering the physical origin of galaxy assembly bias is a critical part of understanding the galaxy -- halo connection \citep{Wechsler2018}. Empirical and semi-empirical models that link galactic growth and halo growth by construction (e.g., \citealt{Hearin2013e, Mutch2013, Moster2018}) have been instrumental in quantifying the impact of assembly bias, but have shed little light on the actual physics.

In principle, hydrodynamical simulations of galaxy formation are well suited to study galaxy assembly bias as they incorporate the relevant physical processes without explicitly assuming a galaxy--halo connection. 
However, previous approaches either suffered from small sample sizes (e.g., \citealt{Feldmann2015, Romano-Diaz2017a}) or focused on galaxies in moderately massive halos ($>10^{11.5}$ $M_\odot$) in large cosmological volumes (e.g., \citealt{Chaves-Montero2016, Matthee2017}).
Consequently, the importance of the halo growth history for low mass galaxies remains to be fully explored.

Fortunately, with the recent advent of large suites of zoom-in galaxy formation simulations it is now feasible to obtain large samples of well-resolved low mass galaxies. In this Letter we use the simulation suite MassiveFIRE \citep{Feldmann2016, Feldmann2017}, which is part of the Feedback in Realistic Environments (FIRE) project\footnote{see \url{http://fire.northwestern.edu}} \citep{Hopkins2014}, to study the impact of the halo growth history on stellar masses and SFRs per unit stellar mass (sSFRs) of low mass, central galaxies at $z\sim{}2$. In order to explore the underlying physical mechanisms, we supplement our hydrodynamical simulations with a simple analytical model that captures the basic behavior of the galaxies in the simulations.

This Letter is organized as follows. We describe our simulation suite and sample selection 
in section \ref{sect:methods}. In section \ref{sect:WhichHaloProperty}, we explore whether or not stellar masses and sSFRs of the galaxies in our sample depend on halo growth rate. We introduce a simple analytic model to discuss our findings in section \ref{sect:Quant}. We explain why halo growth rates affect galaxy masses and sSFRs in section \ref{sect:Explanation} and present our main conclusions in section \ref{sect:Conclusion}.

\section{Methodology and Sample}
\label{sect:methods}

The cosmological, hydrodynamical simulations presented in this Letter are part of the \MassiveFIRE{} suite and are described in detail in \cite{Feldmann2016, Feldmann2017}. In brief, we simulate the formation and growth of galaxies and dark matter halos down to $z\sim{}2$ in 18 distinct zoom-in regions in a (144 comoving Mpc)$^3$ box. Baryonic (star and gas) and high-resolution dark matter particle masses are $3.3\times{}10^4$ $M_\odot$ and $1.7\times{}10^5$ $M_\odot$, respectively. At this resolution, stellar masses, stellar mass profiles, and average SFRs are approximately converged \citep{Hopkins2014, Hopkins2017, Ma2017}.
Gravitational softening lengths are 21 pc for star particles and 143 pc for dark matter. Softening is adaptive for gas particles with a minimum of 9 pc. We created initial conditions with \MUSIC{} \citep{Hahn2011} with $\Omega_{\rm matter}=0.2821$, $\Omega_\Lambda=1-\Omega_{\rm matter}=0.7179$, $\sigma_8=0.817$, $n_{\rm s}=0.9646$, and $H_0=69.7$ km s$^{-1}$ Mpc$^{-1}$ \citep{Hinshaw2013}. All simulations were run with the gravity-hydrodynamics code GIZMO in Pressure-energy Smoothed Particle Hydrodynamics (P-SPH) mode and use the original FIRE physics module\footnote{The recent update of the FIRE module (FIRE-2) does not significantly affect the stellar masses and SFRs of FIRE galaxies in $10^9-10^{12}$ $M_\odot$ halos \citep{Hopkins2017}.}, which includes star formation in gravitationally bound gas clouds, stellar feedback (kinetic and radiative energy and momentum injections, as well as stellar mass loss) from supernova and young stars, chemical enrichment and transport, and gas heating and cooling, see \cite{Hopkins2014} for details. 

\cite{Feldmann2016, Feldmann2017} analyzed the properties of massive ($M_{\rm star}>10^{10}$ $M_\odot$) central galaxies in \MassiveFIRE{}. The present study targets the thousands of low mass central galaxies and their halos scattered throughout the zoom-in regions. Specifically, we select all isolated (i.e., non sub-) halos at $z=2$ with masses in the range $10^{9}-10^{12}$ $M_\odot$ that are not significantly polluted (less than 1\% by mass) by lower resolution dark matter particles with the help of the AMIGA Halo Finder (AHF, 
\citealt{Gill2004}). Throughout this Letter, halo mass refers to the total mass within a virialized overdensity \citep{Bryan1998}. With the help of AHF, we trace the mass growth history of the most massive progenitor of each selected halo using 35 snapshots between $z=10.4$ and $z=2$. We subsequently obtain the detailed halo growth histories for 1543 low mass, central galaxies.
As a consequence of the zoom-in approach, many selected halos reside in the large-scale vicinity of a more massive halo, and halo growth histories of halos in low density environments are potentially underrepresented. To mitigate any arising bias, we explicitly control for the halo growth rate and minimize the use of growth-rate-averaged galaxy properties. Furthermore, we exclude central galaxies that were satellites at any time prior to $z=2$ (`splashback' galaxies). 

The instantaneous growth rate of halos is often highly variable due to mergers, fly-bys, or a temporary tidal truncation. To arrive at a robust estimate of the halo growth rate, we adopt a strategy similar to the one described by \cite{Feldmann2016}, i.e. we fit the assembly history of each given halo with a smooth, parametrized function. Different from \cite{Feldmann2016}, we model the growth rate of individual halos as $\left[dM_{\rm halo}/dt\right]_{\rm fit} = (a + b/\omega) \left[dM_{\rm halo}/d\omega\right]_{\rm avg} d\omega/dt$. Here, $\omega(z)=1.686/D(z)$, $D$ is the linear growth factor normalized to 1 for z = 0, and 
$\left[dM_{\rm halo}/d\omega\right]_{\rm avg}(M, z)$ is the average growth rate of halos of mass $M$ at redshift $z$ following \cite{VandenBosch2014}. Adopting the average growth prescription by \cite{Neistein2006b} results in minor quantitative, but no qualitative, changes of our results. The parameters $a$ and $b$ allow us to model different growth histories, including those with declining halo masses at late times. We compute $a$ and $b$ for all halos in our sample by fitting their growth histories to the parametrized form above via non-linear least squares over the $z=2-6$ redshift range. Throughout this Letter, $\xi\equiv{}a + b/\omega(z=2)$ denotes the $z=2$ growth rate of a halo relative to the average growth rate of $z=2$ halos of the same mass.

Halo concentrations at a redshift $z=2$ are taken from the halo catalogs generated by AHF. The half-mass (5\%-mass) redshift corresponds to the first time that a halo assembled 50\% (5\%) of its maximum mass at $z\geq{}2$.

Galaxy properties (stellar masses, cold gas masses, SFRs, and sSFRs) are obtained from all particles that lie within 10\% of the virial radius of a given halo at $z=2$. Contributions from identified satellite galaxies are excluded. Cold gas refers to gas with $T<1.5\times{}10^4$ K. SFRs are averaged over the past dynamical time of the halo (450 Myr at $z=2$ independent of halo mass), which is somewhat longer (shorter) than the time scales probed by rest-frame ultra-violet (far infrared) light, e.g., \cite{Calzetti2012}. The SHMRs and SFR -- stellar mass relations of $z=2-6$ dwarf galaxies in \MassiveFIRE{} are consistent with previous FIRE literature \citep{Sparre2015, Fitts2017, Ma2017}.

\section{Do galaxy properties depend on the halo growth rate?}
\label{sect:WhichHaloProperty}

\begin{figure*}
\begin{tabular}{c}
\includegraphics[width=170mm]{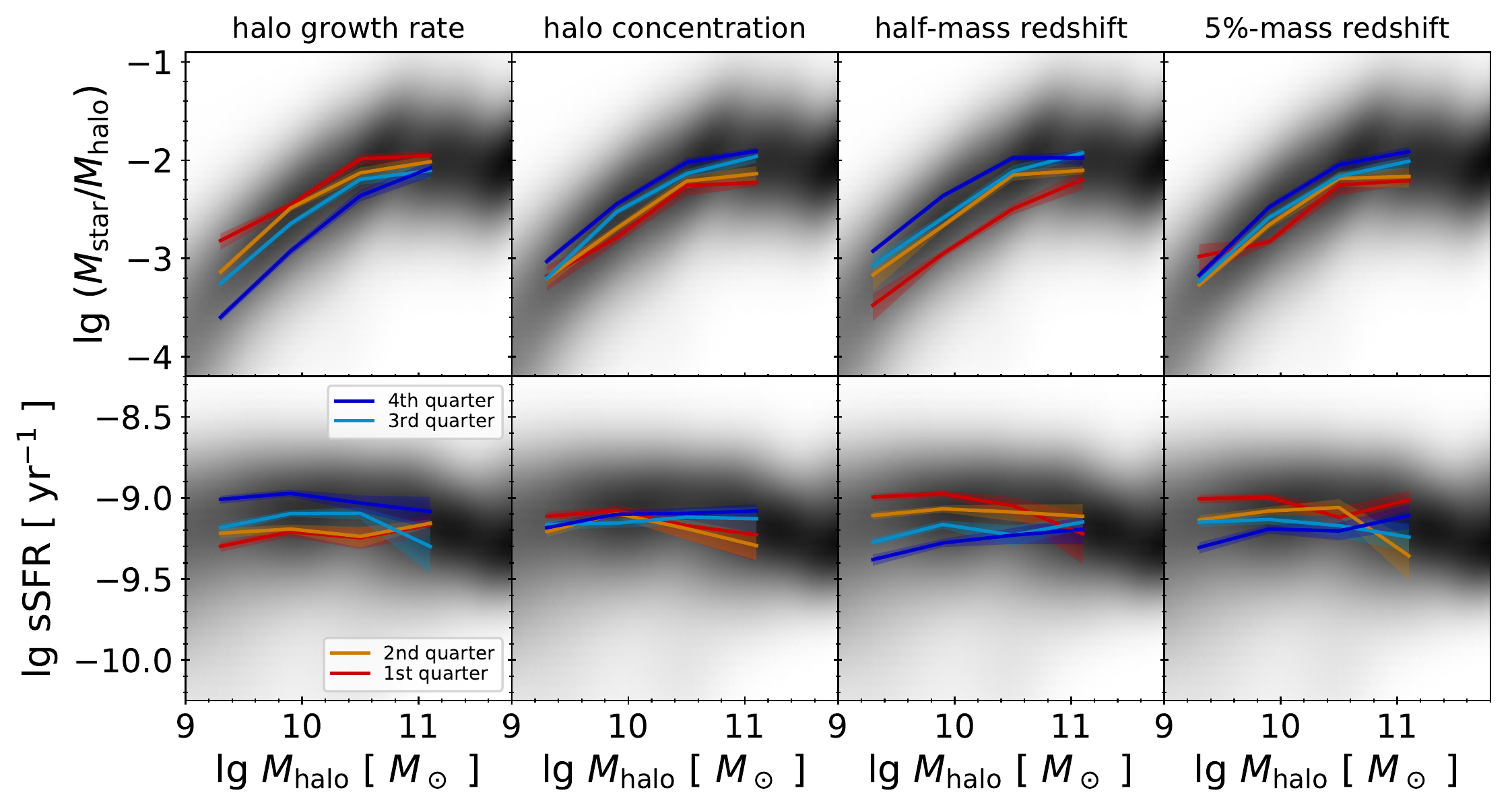}
\end{tabular}
\caption{\emph{At fixed halo mass, stellar masses and sSFRs of central galaxies depend on the halo growth rate.} The gray background in the top row is an estimate of the conditional probability density of the stellar mass fraction, i.e, $p(\lg(M_{\rm star}/M_{\rm halo})\vert{}M_{\rm halo})$, for the 1543
central (i.e., non-satellite) galaxies in our $z=2$ sample of simulated galaxies.
The gray background in the bottom row similarly depicts the conditional probability density of the sSFR, i.e., $p(\lg({\rm sSFR})\vert{}M_{\rm halo})$, for the same galaxies in the sample.
The overall sSFR distribution does not appear to vary strongly with halo mass over the range shown in the figure. Colored lines show the average stellar mass fraction and sSFR in equal-sized bins of various halo properties.
Specifically, the first (fourth) bin contains the 25\% of the galaxies with the lowest (largest) value of the halo growth rate, halo concentration, halo half-mass redshift, and halo 5\%-mass redshift (columns from left to right). Errors are computed via bootstrapping. Results for bins with less than five galaxies are omitted. Stellar masses and stellar mass fractions decrease systematically with increasing halo growth rate, while the sSFRs increase. Stellar masses and sSFRs show similar, albeit in some cases weaker, trends with halo concentration, half-mass redshift, and 5\%-mass redshift.
}
\label{fig:multiPanel}
\end{figure*}

In order to address this question properly, we need to disentangle the effects of halo growth rate and halo mass. In Fig.~\ref{fig:multiPanel} we show basic properties of low mass, central galaxies, their stellar mass fractions and sSFRs, as a function of halo mass. In agreement with previous estimates, stellar mass fractions increase with halo mass while sSFRs are approximately independent of halo mass (e.g., \citealt{Behroozi2013c}). We then split our sample into four equal-sized bins for each of the following halo properties: halo growth rate, halo concentration, half-mass redshift and 5\%-mass redshift. The colored lines in Fig.~\ref{fig:multiPanel} show stellar mass fractions and sSFRs as function of halo mass in each of the four bins and for each of the halo properties.

Fig.~\ref{fig:multiPanel} shows that, at fixed halo mass, halo growth rate correlates with the stellar masses and sSFRs of galaxies. Halos that grow faster at $z\gtrsim{}2$ harbor galaxies with lower stellar masses and higher sSFRs at $z=2$. Furthermore, at fixed halo mass, more concentrated halos and halos with earlier formation time tend to contain galaxies with higher stellar masses and lower sSFRs. However, these latter trends are not always as strong as the scalings with halo growth rate. 
Given that gas cooling and stellar feedback affect the density profile of halos, we expect a stronger dependence on concentration if the (observationally inaccessible) concentration of the same halos in dark-matter-only simulations is used as a predicting variable \citep{Fitts2017}.
While Fig.~\ref{fig:multiPanel} only shows results for $z=2$, the trends are qualitatively similar for any $z=2-6$. Hence, at fixed halo mass, halo growth rate appears to be a fundamental predictor of stellar mass and sSFRs, arguably even more so than halo concentration and halo formation redshift. We stress that our findings apply only to \emph{low mass} halos and may not hold for galaxies in slowly-cooling, massive ($M_{\rm halo}>10^{12}$ $M_\odot$) halos.

\section{Quantifying and modeling the role of the halo growth rate}
\label{sect:Quant}

\begin{figure}
\begin{tabular}{c}
\includegraphics[width=85mm]{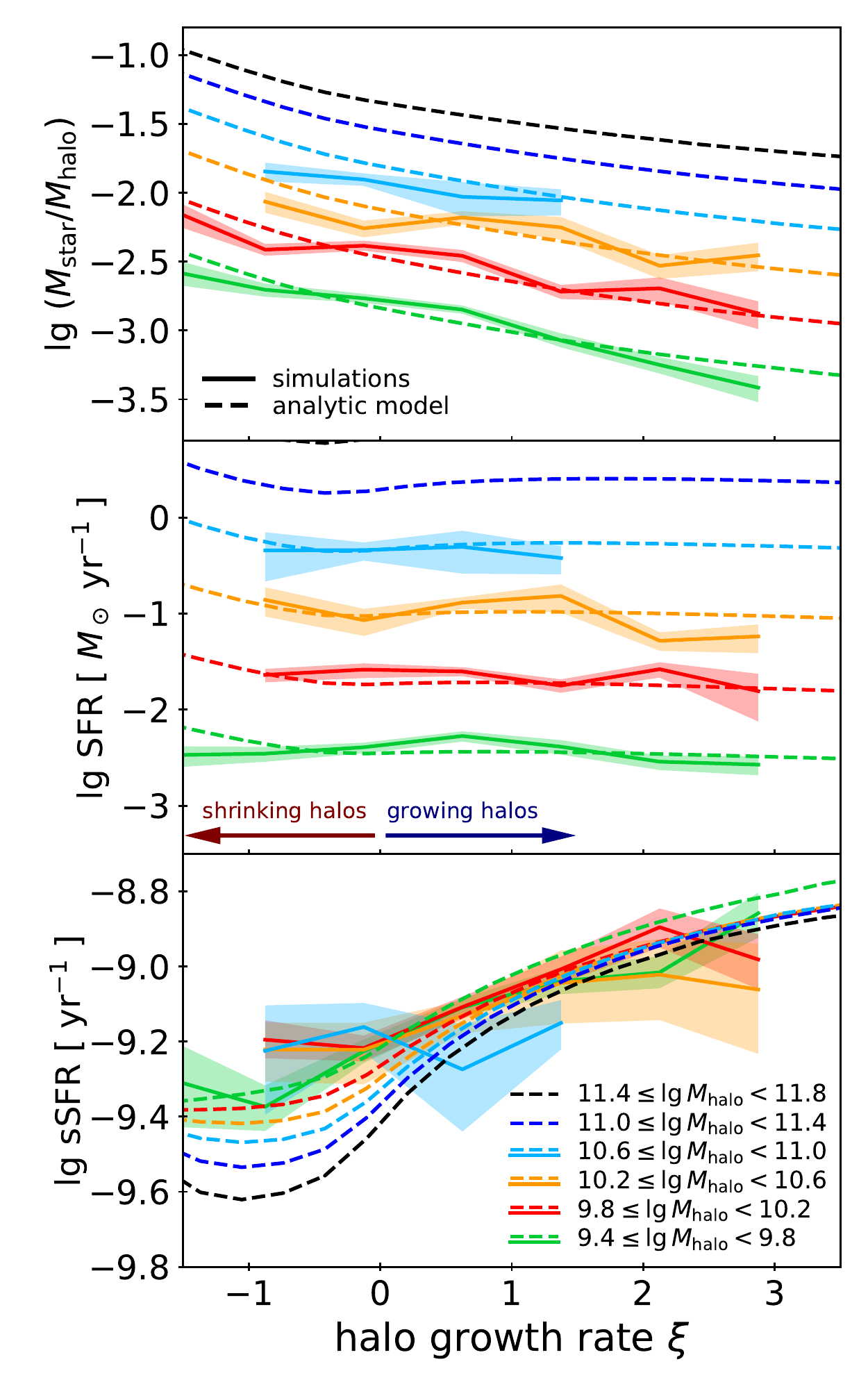}
\end{tabular}
\caption{\emph{A simple analytic model reproduces the dependence of stellar mass and sSFR on halo growth rate.} Solid lines show results for all simulated, central galaxies in our $z=2$ sample. The shaded regions are error bars estimated via bootstrapping. Dashed lines are predictions of a simple analytic model discussed in the text. The x-axis shows the halo growth rate normalized to the typical growth rate for halos of the same mass. 
The analytic model matches the simulations predictions over a large range in halo masses and growth rates. At fixed halo mass, stellar masses and sSFRs vary systematically with halo growth rate.}
\label{fig:dependenceOnAccretion}
\end{figure}

As argued in the previous section, stellar masses and sSFRs vary with halo growth rate in a systematic and monotonic way. Fig.~\ref{fig:dependenceOnAccretion} sheds additional light on this result. Here, we show stellar mass fractions, SFRs, and sSFRs for the galaxies in our sample as a function of halo growth rate. Interestingly, while stellar masses depend on both halo growth rate and mass, SFRs are essentially independent of halo growth rate at fixed halo mass. In contrast, sSFRs scale with halo growth rate but only weakly with halo mass over the $M_{\rm halo}=10^{9.4}-10^{11}$ $M_\odot$ range.

We fit stellar mass, SFR, and sSFR as functions of halo accretion rate and halo (or stellar) mass with the help of the publicly available VGAM R package \citep{Yee2015}. 
In order to deal properly with galaxies with zero star formation, we use zero-inflated negative binomials as family functions when fitting count-based SFRs and sSFRs \citep{Feldmann2017a}. The average SFRs and sSFRs reported below exclude non-star forming galaxies corresponding to the zero-inflated component. For consistency, we also use negative binomials as family functions when fitting stellar masses (interpreted as count data using the number of stellar particles in the galaxy).

Average stellar masses, SFRs, and sSFRs of low mass, central galaxies show the following scaling behavior:
{\small
\begin{eqnarray*}
\lg \langle{}M_{\rm star}\rangle{} &=& (-0.220\pm0.011)\xi + (1.94\pm0.02) \lg M_{\rm halo} + {\rm const},\\
\lg \langle{}{\rm SFR}\rangle{} &=& (-0.047\pm0.015)\xi + (1.91\pm0.03) \lg M_{\rm halo} + {\rm const}, \\
\lg \langle{}{\rm SFR}\rangle{} &=& (0.098\pm0.011)\xi + (1.04\pm0.02) \lg M_{\rm star} + {\rm const}, \\
\lg \langle{}{\rm sSFR}\rangle{} &=& (0.093\pm0.010)\xi + (0.086\pm0.024) \lg M_{\rm halo} + {\rm const}.
\end{eqnarray*}
}

The scaling of the average sSFRs with halo growth rate $\xi$ is somewhat shallower than expected from the scaling exponents for average stellar masses and SFRs at fixed halo mass. Positive correlations between (past-averaged) SFRs and current stellar masses as well as the considerable overall scatter, result in a reduced correlation for the sSFR.

In order to explore the origin of the dependence of stellar masses and sSFRs on halo growth rate, we make use of a simple analytic model (Feldmann in prep). It is a one-zone model with two free parameter functions, each potentially depending on halo mass and cosmic time but not explicitly on any other halo or galaxy property. The first free parameter is the cold gas depletion time $t_{\rm dep}$, which specifies how quickly the galaxy converts cold gas into stars (${\rm SFR}=M_{\rm cgas}/t_{\rm dep}$). The second parameter, $f_{\rm cbar}$, is the amount of cold baryons (stars and cold gas) in the galaxy relative to the amount of baryons in the halo ($M_{\rm star}+M_{\rm cgas}=f_{\rm cbar}f_{\rm b}M_{\rm halo}$). Stellar masses of low mass galaxies are predominantly build by in-situ star formation (e.g., \citealt{Behroozi2013c, Angles-Alcazar2017}), and, hence, we will ignore galaxy mergers in the following, simplified discussion. 

Combining the two previous equations, we arrive at an ordinary differential equation for the stellar mass:
$dM_{\rm star}/dt/(1-R) = {\rm SFR} = (f_{\rm cbar}f_{\rm b}M_{\rm halo} - M_{\rm star})/t_{\rm dep}$, where $R$ is the gas return fraction in the instantaneous recycling approximation. For a given halo growth history, $M_{\rm halo}(t)$, the parameters $t_{\rm dep}$ and $f_{\rm cbar}$ can be interpreted as functions of only time and the differential equation above can be solved analytically to obtain $M_{\rm star}(t)$ and ${\rm SFR}(t)$. This simple model predicts stellar mass and SFR histories for any halo history provided that the functions $t_{\rm dep}$ and $f_{\rm cbar}$ are known.

We adopt simple parametrized forms for $t_{\rm dep}$ and $f_{\rm cbar}$ and determine the parameters from a maximum a posteriori estimation with non-informative priors based on our $z=2$ galaxy sample. This calibration constrains how $t_{\rm dep}$ and $f_{\rm cbar}$ depend on halo mass and cosmic time. Importantly, the model does not introduce an explicit dependence on halo growth rate. Our model calibration results in cold gas depletion times of about 1 Gyr at $z=2$, only weakly dependent on halo mass, but moderately increasing with decreasing redshift, in qualitative agreement with observations \citep{Genzel2015}. The cold baryon fraction varies with halo mass (but is assumed to be redshift independent), peaking at $\sim{}20\%$ of the universal baryon fraction in $M_{\rm halo}\sim{}10^{11-12}$ $M_\odot$ halos.

Dashed lines in Fig.~\ref{fig:dependenceOnAccretion} show the model predictions with $t_{\rm dep}$ and $f_{\rm cbar}$ calibrated as described above. The simple model matches the results of our hydrodynamical simulations quite well. In particular, it shows that sSFRs of low mass, central galaxies correlate with the growth rate of their halos, while stellar masses anti-correlate with halo growth rate, and SFRs scale primarily with halo mass but not with halo growth rate.

\section{Why do galaxy properties depend on halo growth rate?}
\label{sect:Explanation}

\begin{figure}
\begin{tabular}{c}
\includegraphics[width=85mm]{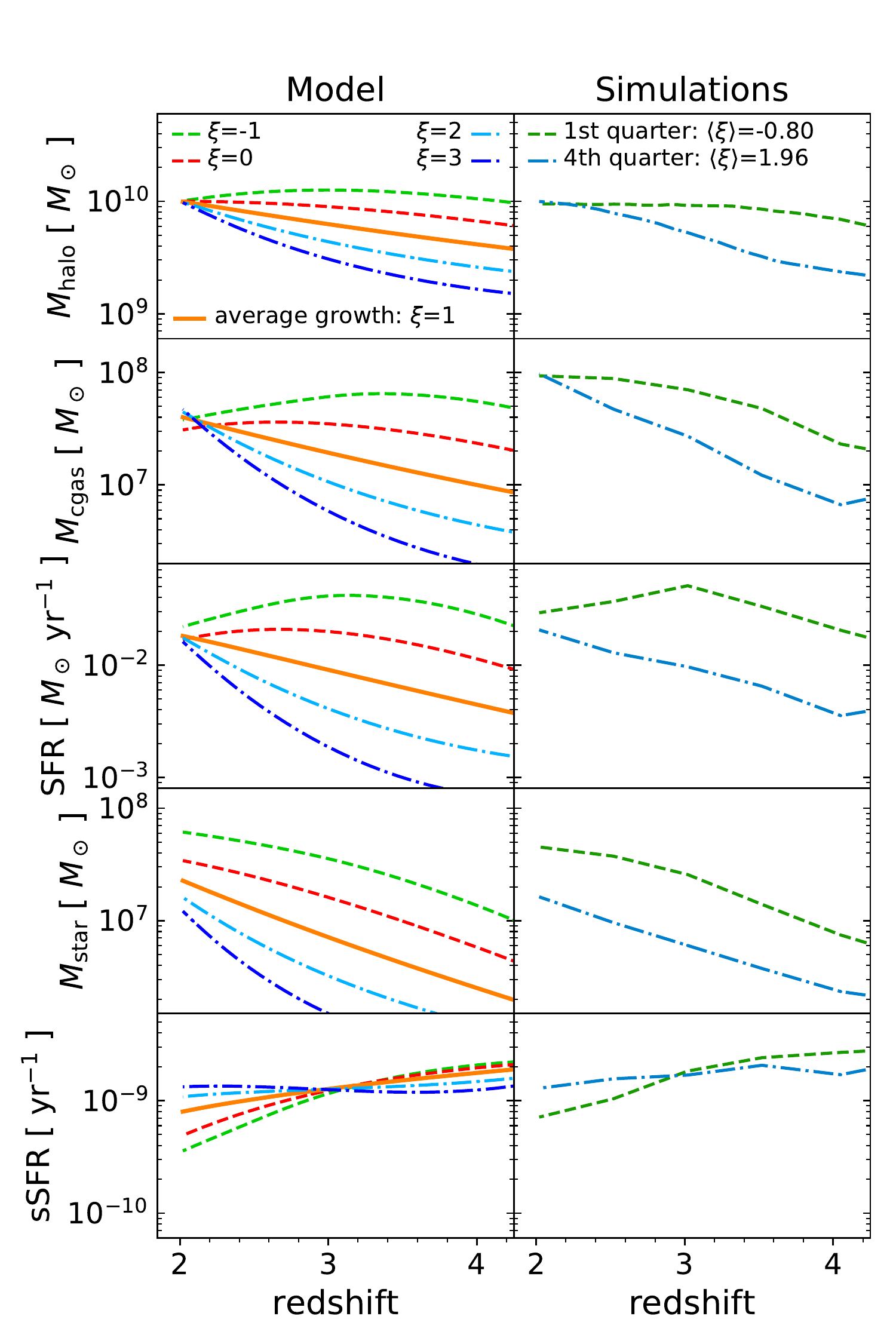}
\end{tabular}
\caption{\emph{Among halos of the same mass, halos that are growing faster at $z=2$ had lower gas masses and SFRs in the past, resulting in lower stellar masses and higher sSFRs at $z=2$.} The plots in the left column show the evolution of halo mass (top panel) and various galaxy properties (cold gas mass, instantaneous SFR, stellar mass, and instantaneous sSFRs) as predicted by the analytic model (see the text) for halos with a $z=2$ mass of $10^{10}M_\odot$ and different halo growth rates. The right column shows the same properties for our simulation sample. Specifically, we show the average properties of all halos with $9.8\leq{}\lg{}M_{\rm halo}(z=2)/M_\odot<10.2$ that lie within either the top or bottom quarter of the $\xi$ distribution.}
\label{fig:explanation}
\end{figure}

Fig.~\ref{fig:explanation} offers a straightforward explanation for the results presented in this Letter.
Halos that grow more quickly near $z=2$ were naturally less massive at earlier times than halos that grow more slowly near $z=2$. At $M_{\rm halo}<10^{12}$ $M_\odot$, less massive halos contain less cold gas. The scaling is super-linear at low halo masses due to efficient gas ejections via stellar feedback (e.g., \citealt{Muratov2015, Angles-Alcazar2017}) and decreased gas accretion resulting from photo-ionization suppression (e.g., \citealt{Okamoto2008a, Faucher-Giguere2011, Fitts2017}).
Lower cold gas masses at earlier times imply lower SFRs in the past. Hence, the stellar mass (the integral of the SFR) near $z=2$ tends to be lower in more quickly growing halos. Consequently, sSFRs at $z=2$ will be higher for halos that grow more quickly near $z=2$, and they will be lower in more slowly growing halos. 

Our results provide a physical explanation for the ansatz of conditional abundance matching (CAM) models which assume that, at a given halo mass, sSFRs of galaxies depend on halo growth history \citep{Hearin2013e}. Given the findings presented in section \ref{sect:WhichHaloProperty}, we propose that CAMs also explore the use of halo growth rate as an alternative measure of halo growth history in addition to halo concentration and formation time.

The SFR at $z=2$ is only weakly dependent on halo growth rate in halos of the same mass. The explanation is rather simple in the context of our model. Halos with the same mass at $z=2$ have (by virtue of our model) also the same mass of cold baryons and the same cold gas depletion time. As cold gas, and not stars, makes up most of the cold baryons in low mass galaxies, halos of the same mass have similar cold gas masses and, hence, SFRs. 
In contrast, the cold gas-to-stellar mass ratio, $\mu=t_{\rm dep}\,{\rm sSFR}$, varies with halo growth rate. Specifically, at fixed halo mass, more quickly growing halos will have a larger $\mu$ than more slowly growing halos.

For low and negative halo growth rates the picture becomes slightly more complex. On the one hand, the much lower $\mu$ in halos with very low growth rates results in lower cold gas masses and SFRs. On the other hand, galaxies in shrinking halos can have cold baryon masses that are much larger than those of growing halos, even at fixed halo mass, because a reduction in halo mass (via e.g., tidal stripping) does not necessarily result in a similar reduction in cold baryonic mass. Hence, the model predicts that galaxies in shrinking halos have larger absolute cold gas masses and SFRs than galaxies in growing halos of the same final mass, see Fig.~\ref{fig:dependenceOnAccretion}.

\section{Conclusions}
\label{sect:Conclusion}

The dependence of stellar masses and sSFRs on halo growth rate may contribute to the observed scatter in galaxy scaling relations. 
For instance, \cite{Rodriguez-Puebla2016} suggested that the scatter in SFR at fixed stellar mass could be explained this way. Observations constrain this scatter to be 0.2-0.4 dex over a large range of masses and redshifts (e.g., \citealt{Schreiber2015b}). 

Given the scalings presented in section \ref{sect:Quant}, we expect that SFRs change by $\sim{}0.1$ dex per unit change of $\xi$ \emph{at fixed stellar mass}. We measured the scatter of the halo growth rate both in our sample (at $z=2$) and in collisionless, cosmological volume simulations (at $z=2$ and $z=0$) finding $\sigma_\xi\sim{}1$ for halos in the $\sim{}10^{9-11}$ $M_\odot$ mass range. Hence, it appears that variations in halo growth rate cannot account for more than $\sim{}0.1$ dex of scatter in the star forming sequence. In fact, in our simulations, much of the scatter of the star forming sequence arises from galaxy related processes that operate on time-scales that are shorter than the dynamical time of the halo, such as fluctuations of the gas accretion rate onto galaxies and frequent, but short-lived, bursts of star formation \citep{Sparre2015, Feldmann2017, Faucher-Giguere2018}.

Combining $\sigma_\xi\sim{}1$ with the scaling of the average stellar mass with $\xi$, see section \ref{sect:Quant}, we predict $\sim0.22$ dex of scatter in stellar mass at fixed halo mass. This predicted scatter matches approximately the scatter in our sample at $z=2$ ($\sim{}0.2 $ dex at $M_{\rm halo}=10^{11}$ $M_\odot$ and $\sim{}0.28$ dex at $M_{\rm halo}=10^{10.25}$ $M_\odot$). The overall amount of scatter and its increase with decreasing halo mass in dwarf galaxy halos is consistent with results of previous hydrodynamical simulations \citep{Wang2015, Pillepich2018, Wechsler2018}. Empirically, the scatter evolves only weakly with redshift \citep{Behroozi2018a}, reaching $\sim{}0.18-0.2$ dex in moderately massive to very massive halos at $z=0$ (from $M_{\rm halo}\sim{}10^{12}$ $M_\odot$ to galaxy clusters, e.g., \citealt{Zu2015a, Kravtsov2018}). Variations in the halo growth histories may thus explain a large fraction of the scatter in stellar mass at fixed halo mass in low mass, central galaxies.

\acknowledgments
The authors would like to thank the referee for valuable comments that improved the quality of the manuscript. The authors also thank Onur \c{C}atmabacak for providing halo catalogs for two dark-matter-only simulations used in this Letter. RF acknowledges financial support from the Swiss National Science Foundation (Grant No. 157591). CAFG was supported by NSF through grants AST-1412836, AST-1517491,ÊAST-1715216, and CAREER award AST-1652522, by NASA through grant NNX15AB22G, and by a Cottrell Scholar Award from the Research Corporation for Science Advancement. DK was supported by NSF grant AST-1715101 and the Cottrell Scholar Award from the Research Corporation for Science Advancement. This research was supported in part by the National Science Foundation under grant No. NSF PHY-1748958. Simulations were run with resources provided by the NASA High-End Computing (HEC) Program through the NASA Advanced Supercomputing (NAS) Division at Ames Research Center, proposal SMD-14-5492. Additional computing support was provided by HEC allocations SMD-14-5189, SMD-15-5950, by NSF XSEDE allocations AST120025, AST150045, by allocations s697, s698 at the Swiss National Supercomputing center, and by S3IT resources at the University of Zurich. This work made extensive use of the NASA Astrophysics Data System and arXiv.org preprint server.\\
\\[0cm]

\bibliographystyle{aasjournal}

 \end{document}